\documentclass[
aps,
pra,
twocolumn,
floatfix,
]{revtex4-2} 
\usepackage{amsmath,amsthm,amssymb,epsfig,graphicx,mathrsfs,amsfonts,dsfont,bbm}
\usepackage{pict2e}
\usepackage[percent]{overpic}
\usepackage{color}
\usepackage{listings}
\usepackage[caption=false]{subfig}
\usepackage[toc,title,titletoc,header]{appendix}
\usepackage{color}
\usepackage{dcolumn}
\usepackage{bm}
\usepackage{hyperref}
\hypersetup{
    citecolor=magenta,
    colorlinks=true,
    linkcolor=blue,
    filecolor=green,      
    urlcolor=cyan,
}
\usepackage[capitalise]{cleveref}
\usepackage{enumitem}
\setlist[enumerate]{leftmargin=10mm, label=\alph*)}
\setlist[itemize]{leftmargin=12pt}
\usepackage{mathtools}
\usepackage{tikz}
\usepackage{braket}
\usepackage{physics}
\usepackage{luatex85,qcircuit}
\usepackage{blkarray}
\usepackage[linesnumbered,ruled,vlined,algosection]{algorithm2e}

\SetCommentSty{mycommfont}

\theoremstyle{plain}

\theoremstyle{definition}
\newtheorem{definition}{Definition}
\newtheorem*{problem*}{Problem} 
\newtheorem{problem}{Problem}

\newcommand{\hilbertspace}{\mathcal{H}}
\newcommand{\bigO}{\mathcal{O}}

\newcommand{\realnumber}{\mathbb{R}}

\newcommand{\identity}{\mathds{1}}

\newcommand{\ii}{\textup{i}}

\newcommand{\expectation}{\mathbb{E}}

\DeclareMathOperator{\poly}{\textup{poly}}

\newcommand{\T}{\intercal}

\newcommand\vartextvisiblespace[1][.5em]{%
  \makebox[#1]{%
    \kern.07em
    \vrule height.3ex
    \hrulefill
    \vrule height.3ex
    \kern.07em
  }%
}

\newcommand{\vbx}{\vb{x}}

\newcommand{\vbw}{\vb{w}}

\usepackage{stmaryrd}

\newcommand{\kernel}{k}
\newcommand{\ew}{W}
\newcommand{\ob}{O}
\newcommand{\pob}{O}
\newcommand{\dm}{\rho}
\newcommand{\ghz}{\text{GHZ}}

\newcommand{\ml}{\text{ml}}
\newcommand{\bi}{\text{bi}}
\newcommand{\cs}{\text{cs}}

\newcommand{\perm}{\textup{Perm}}
\newcommand{\target}{\textup{tar}}
\newcommand{\prepare}{\textup{pre}}
\newcommand{\entangled}{\textsc{entangled}}
\newcommand{\separable}{\textsc{separable}}
\newcommand{\noise}{\text{noise}}

\newcommand{\chsh}{\textup{CHSH}}
\newcommand{\bellineq}{\textup{Bell}}

\newcommand{\separableset}{\mathcal{S}}

\newcommand{\px}{X}

\newcommand{\pz}{Z}

\newcommand{\bmsigma}{\bm{\sigma}}

\newcommand{\U}{U}

\newcommand{\ppartition}{\mathcal{P}}

\makeatletter
\def\l@subsubsection#1#2{}
\makeatother

\begin{document}
\title{Towards efficient and generic entanglement detection by machine learning}
\author{Jue Xu}
\email{juexu@cs.umd.edu}
\author{Qi Zhao}
\email{zhaoqi@cs.hku.hk}
\affiliation{
QICI Quantum Information and Computation Initiative, Department of Computer Science,
The University of Hong Kong, Pokfulam Road, Hong Kong}
\date{\today}
\begin{abstract}
	Detection of entanglement is an indispensable step to practical quantum computation and communication. 
	Compared with the conventional entanglement witness method based on fidelity, we propose a flexible, machine learning assisted entanglement detection protocol that is robust to different types of noises and sample efficient.
	In this protocol, an entanglement classifier for a generic entangled state is obtained by training a classical machine learning model with a synthetic dataset.
	The dataset contains classical features of two types of states and their labels (either entangled or separable). 
	The classical features of a state, which are expectation values of a set of $k$-local Pauli observables, are estimated sample-efficiently by the classical shadow method. 
	In the numerical simulation, our classifier can detect the entanglement of 4-qubit GHZ states with coherent noise and W states mixed with large white noise, with high accuracy.
\end{abstract}

\maketitle

\section{Introduction}
Entanglement \cite{horodeckiQuantumEntanglement2009} is the key ingredient of quantum teleportation \cite{bennettTeleportingUnknownQuantum1993}, quantum cryptography \cite{ekertQuantumCryptographyBased1991}, quantum computation \cite{briegelMeasurementbasedQuantumComputation2009}, and quantum metrology \cite{giovannettiQuantumenhancedMeasurementsBeating2004}.
However, decoherence and imperfections are inevitable in real-world devices, which means the interaction between a quantum system and a classical environment would significantly affect entanglement quality and diminish quantum advantage in applications. 
For practical purposes, it is essential to detect entanglement in certain quantum physical systems.
This problem has been widely studied \cite{guhneEntanglementDetection2009}, but still far from being perfectly solved. 

Quantum tomography, as one of the most widely used certification methods, can provide the full density matrix of the prepared state. However, even given the tomography results, it is computationally intractable to determine whether the state is entangled by classical \cite{gurvitsClassicalDeterministicComplexity2003} or quantum computation \cite{gutoskiQuantumInteractiveProofs2015}.
Not alone, the sample complexity of quantum tomography
grows exponentially with dimension \cite{haahSampleoptimalTomographyQuantum2017,odonnellEfficientQuantumTomography2016}.
Thus, a more realistic scenario is entanglement witnesses that can determine whether a prepared state is entangled or not with the prior knowledge of the state. 
This task for many entangled states of practical interest can be efficiently solved by measuring a few observables \cite{bourennaneWitnessingMultipartiteEntanglement2004,tothDetectingGenuineMultipartite2005,tothEntanglementDetectionStabilizer2005}.
Though attempts such as \cite{guhneNonlinearEntanglementWitnesses2006,zhouEntanglementDetectionCoherent2020} have been made to enhance robustness to noise, entanglement witnesses will also fail when there is a lot of noise or unexpected types of noise in practice 
\cite{weilenmannEntanglementDetectionMeasuring2020}.
Moreover, for given witnesses, it is also generally challenging to reduce the measurement efforts (sample complexity), especially for non-stabilizer states \cite{zhangEfficientEntanglementGeneration2021}. 

The goal of this paper is to find an efficient and generic way to detect the entanglement of many-body quantum states. 
Machine learning (ML) is a powerful tool for such a purpose. 
Many ML techniques including both classical and quantum machine learning models have been proposed for classification tasks in physics, such as the classification of phases and prediction of ground states \cite{carrasquillaMachineLearningPhases2017,congQuantumConvolutionalNeural2019,huangProvablyEfficientMachine2022}.
Entanglement detection as a typical classification problem has been studied by ML techniques, such as determining separability by Neural Network (NN) \cite{luSeparabilityEntanglementClassifierMachine2018,maTransformingBellInequalities2018} and deriving generic entanglement witnesses by Support Vector Machine (SVM) \cite{zhuMachineLearningDerivedEntanglement2021,vintskevichClassificationFourqubitEntangled2022}. 
Nevertheless, these prior machine learning assisted methods 
only explore white noise robustness without considering other types of noises that happened in experiments.
And the sample efficiency of experimental implementation for these ML-derived classifiers has not been discussed.

In this work, an ML classifier is obtained by training SVM with a synthetic dataset on a classical computer.
The dataset consists of two types of states, one is a set of certain target entangled states subject to randomly sampled noise, and the other is a set of randomly sampled separable states with the given partitions.
To increase the feasibility in experiments, each state is characterized by its expectation values of Pauli observables, called classical features.
Within the framework of SVM, classification capability can be boosted by nonlinear kernel method and unimportant features can be eliminated programmatically.
Furthermore, we restrict the Pauli observables to $k$-local such that classical features can be estimated with a smaller sample complexity via the classical shadow method \cite{huangPredictingManyProperties2020}.
In the numeric simulation of 4-qubit GHZ state and W state, the kernel SVM classifier exhibits better robustness to white noise than conventional fidelity witnesses and also robust to coherent noise which is more realistic in experiments but not widely studied. 
And the derandomized classical shadow method outperforms other schemes for estimating many $k$-local observables (features).

This paper is organized as follows: in \cref{sec:preliminaries}, we briefly present necessary definitions of multipartite entanglement, related entanglement detection problems, and mainstream methods for these problems;
\cref{sec:protocol} demonstrates our end-to-end protocol including two parts: learning an entanglement witness for a generic state from synthetic data and efficient estimation of classical features of states from experiments;
at last, numerical simulation results are discussed in \cref{sec:numerical_simulation}.

\section{Preliminaries}\label{sec:preliminaries}

\subsection{Multipartite entanglement}

Large-scale entanglement involving multiple particles may be the main resource for quantum advantages in quantum computation and communication.
Roughly, we say a quantum state $\dm$ of $n$ subsystems is \emph{entangled} if it is not fully separable,
i.e., the state cannot be written as the tensor product of all subsystems as $\dm=\rho_1\otimes\cdots\otimes\rho_n$.
Clearly, the simple statement `the state is entangled' would allow only two of the particles are entangled while the rest is in a product state, which is very weak entanglement.
So, the more interesting entanglement property is bipartite separability:
\begin{definition}[bi-separable]\label{def:bipartite_separable}
	A pure state $\ket{\psi}$ is bipartite separable (bi-separable) if and only if it can be written as a tensor product form 
	$\ket{\psi}_{\bi}^{\ppartition} = \ket{\phi_{A}}\otimes\ket{\phi_{B}}$ with some bi-partition $\ppartition=\qty{A,B\equiv \bar{A}}$. 
	A mixed state $\dm$ is bi-separable if and only if it can be written as a convex combination of pure bi-separable states, i.e.,
	$\dm_{\bi}=\sum_i p_i \op{\psi_i}_\bi^{\ppartition_i}$ 
	($\ppartition_i$ can be different partitions)
	with a probability distribution $\qty{p_i}$.
	The set of all bi-separable states is denoted as $\separableset_\bi$.
\end{definition}
\begin{definition}[GME]\label{def:gme}
	On the contrary, if a state $\dm\notin \separableset_\bi$,
	it possesses genuine multipartite entanglement (GME).
\end{definition}

GME implies that all subsystems are indeed entangled with each other,
so it is the strongest form of entanglement. 
Whereas, there is another restricted way for generalizing bi-separability to mixed states: 
if it is a mixing of pure bi-separable states with the same partition $\ppartition_2$, 
and we denote the state set as $\separableset_{\bi}^{\ppartition_2}$. 
It is practically interesting to study entanglement under the certain partition,
because it naturally indicates the quantum information processing capabilities among a real geometric configuration.
We have a formal definition for entanglement concerning partitions:
\begin{definition}[full entanglement]\label{def:full_entanglement}
	A state $\dm$ possesses full entanglement
	if it is outside of the separable state set $\separableset_{\bi}^{\ppartition_2}$ for any partition,
	that is, $\forall \ppartition_2 = \qty{A,\bar{A}},\dm \notin \separableset_\bi^{\ppartition_2}$.
\end{definition}
For a state with full entanglement, it is possible to prepare it by mixing bi-separable states with different bipartitions,
so full entanglement is weaker than \nameref{def:gme} but still useful in practice.

\subsection{Entanglement detection}
After introducing the definitions of entanglement, 
the next basic question is how to determine the entanglement of a state efficiently.
Despite clear definitions, it is a highly non-trivial question for a general state.
For a general review on this subject, we refer readers to \cite{guhneEntanglementDetection2009}.
One of the most widely studied problems in this area is bi-separability.
\begin{problem}[separability]\label{prm:separability}
	Given a density matrix 
	\footnote{
		A quantum (mixed) state $\dm$ can be represented by a density matrix which is a Hermitian, positive semidefinite operator (matrix) of trace one. If the rank of $\dm$ is 1, then the state is a pure state.
	}
	$\dm$, to determine if it is \nameref{def:bipartite_separable} (in $\separableset_{\bi}$).
\end{problem}

It is not hard to prove that if a state is bi-separable regarding $\ppartition=\qty{A,B}$, then it must have positive partial transpose (PPT),
i.e., 
the partially transposed (PT) 
\footnote{
	The partial transpose (PT) operation acting on subsystem $A$ is defined as
	$\op{k_A,k_B}{l_A,l_B}^{\T_A} := \op{l_A,k_B}{k_A,l_B}$
	where $\qty{\ket{k_A,k_B}}$ is a product basis of the joint system $\hilbertspace_{AB}$.
}
density matrix $\dm_{AB}^{\T_A}$ is positive, semidefinite \footnote{A matrix (operator) is positive, semidefinite (PSD) if all its eigenvalues are non-negative.} \cite{peresSeparabilityCriterionDensity1996,horodeckiSeparabilityMixedStates1996}.
By contrapositive, we have a sufficient condition for (bipartite) entanglement, that is
	if the smallest eigenvalue of partial transpose $\dm_{AB}^{\T_A}$ is negative (NPT), then the state is entangled (cannot be bi-separable with $\ppartition=\qty{A,B}$).
We should mention that the PPT criterion is a necessary and sufficient condition for \nameref{prm:separability} only when the system dimension is low ($d_A d_B \le 6$ where $d_A$ and $d_B$ are the dimensions of two bipartite subsystems respectively) \cite{horodeckiSeparabilityMixedStates1996}.
Therefore, no general solution for the separability problem is known.
Then, a natural question is whether it is possible to solve separability approximately.
By relaxing the definition (promise a gap between two types of states), a reformulation of separability in the theoretic computer science language is
\begin{problem}[Weak membership problem for separability]\label{prm:weak_membership problem_for_separability}
	Given a density matrix $\dm$ with the promise that either (i) $\dm\in \separableset_{\bi}$ or (ii) $\norm{\dm-\dm_{\bi}}\ge \epsilon$ with certain norm, decide which is the case.
\end{problem}
Unfortunately, even if we are given the complete information about a state and promised a gap (error tolerance $\epsilon$), it is still hard to determine separability approximately by classical computation.
\nameref{prm:weak_membership problem_for_separability} 
is NP-Hard for $\epsilon=1/\poly(d_A,d_B)$ with respect to Euclidean norm and trace norm \footnote{
	The Euclidean norm of a matrix $A$ is defined as $\norm{A}_2:=\sqrt{\Tr(A^\dagger A)}$.
	The trace norm of $A$ is defined as $\norm{A}_{\Tr}\equiv\norm{A}_{1}:=\Tr(\abs{A})\equiv\Tr(\sqrt{A^\dagger A})$.
	Correspondingly, trace distance between two density matrices is $d_{\tr}(\dm,\dm') : = \frac{1}{2} \norm{\dm-\dm'}_1$.
} \cite{gurvitsClassicalDeterministicComplexity2003} \cite{gharibianStrongNPHardnessQuantum2009},
while there exists a quasipolynomial-time algorithm with respect to certain norm \cite{brandaoQuasipolynomialtimeAlgorithmQuantum2011}.
A notable numeric method is the powerful criteria called $k$-symmetric extension hierarchy based on SDP \cite{dohertyCompleteFamilySeparability2004} \cite{ioannouComputationalComplexityQuantum2007} \cite{navascuesPowerSymmetricExtensions2009}, 
which also becomes computationally intractable with growing $k$.
The quantum hardness of a series of related separability testing problems were studied in the framework of quantum interactive proofs \cite{gutoskiQuantumInteractiveProofs2015}.
Nevertheless, these hardness results do not rule out the possibility to solve it efficiently with a stronger promise (approximation) or by machine learning (heuristic) techniques powered by data.

\subsubsection{Entanglement witness based on fidelity}\label{sec:entanglement_witness}
A (realistic) variant of \nameref{prm:separability} is how to determine \nameref{def:bipartite_separable} given copies of an unknown state (from experiments) rather than its full density matrix.
In this case, the sample complexity should be considered besides computational complexity.
Since the input to this problem is quantum data (states), directly estimating spectrum or entanglement monotone functions of the reduced density matrix $\rho_A:=\Tr_B(\rho_{AB})$ \cite{ekertDirectEstimationsLinear2002} \cite{horodeckiDirectDetectionQuantum2002} \cite{johriEntanglementSpectroscopyQuantum2017}, e.g., purity, negativity, and entanglement entropy, by quantum measurement and circuits \cite{wang16qubitIBMUniversal2018} \cite{quekMultivariateTraceEstimation2022} is a good option (without fully recovering density matrices).
However, this line of work does not provide capability beyond theoretical complexity bounds (though usually efficient for the one-side test).
The problem we study here is another variant:
\begin{problem}[entanglement detection with prior knowledge]\label{prm:entanglement_detection}
	Given copies of an unknown state $\dm$ (from experiments) that is promised either (i) $\dm\in\separableset_{\bi}$
	or (ii) in `proximity' of a target $\ket{\psi_{\target}}$,
	determine which is the case.
\end{problem}
The typical scenario for this problem is to prepare a pure entangled state $\ket{\psi_\target}$ in experiments and would like to detect (verify) it as true multipartite entangled. 
While the preparation is not perfect, 
it is reasonable to assume that the prepared mixed state $\dm_{\prepare}$ is in the proximity of the target state,
that is, $\ket{\psi_{\target}}$ undergoes noise channels restricted to white noise and local rotation (unitary).
This problem is supposed to be solved efficiently because we have a much stronger promise than the separability problem.
The usual method for it is constructing an observable $W$ called \emph{entanglement witness} such that
\begin{equation}
	\Tr(\ew\dm_{\bi}) \ge 0  \text{ and }
	\Tr(\ew\op{\psi_{\target}}) < 0 
	\label{eq:witness}
\end{equation}
which means that the witness $W$ has a positive expectation value on all separable states. 
Hence, a negative expectation value implies the presence of entanglement (GME).
It can be proved, for every entangled state, a witness can always be constructed,
but no entanglement witness works for all entangled states \cite{heinosaariMathematicalLanguageQuantum2011}.
So, entanglement witness only provides a one-side test for separability.
For instance, the Bell (CHSH) inequalities originally proposed to rule out local hidden variable models,
can be regarded as an entanglement witness for many 2-qubit entangled states \cite{terhalBellInequalitiesSeparability2000}.
A Bell inequality can be considered as a linear combination of Pauli observables $\ew_{\bellineq}:=\vb{w}_{\bellineq}\cdot\vb{\pob}_{\bellineq}$
such that only entangled states $\dm$ have $\abs{\Tr(\dm\ew_{\bellineq})}$ greater than a threshold
\footnote{
	The Bell (CHSH) inequality (witness):
	$\vb{\pob}_{\chsh}=\qty(\identity, a b, a b', a' b, a' b' )$ with 
	$a = \pz, a' = \px, b = (\px-\pz)/\sqrt{2}, b = (\px+\pz)/\sqrt{2}$
	and $\vb{w}_{\chsh} = \qty(\pm 2, 1, -1, 1, 1)$
}.

While various methods for constructing an entanglement witness exist, the most common one is based on the fidelity between a prepared state $\dm_{\prepare}$ to the target (pure entangled) state $\ket{\psi_{\target}}$
\begin{equation}
	\ew_{\psi} = \alpha\identity - \op{\psi_\target} 
	\label{eq:entanglement_witness}
\end{equation}
where $\alpha = \max_{\dm_{\bi}} \Tr(\dm_{\bi}\op{\psi_{\target}})$ is the maximal fidelity between separable states and the target entangled state such that for every separable state $\Tr(\dm_{\bi}\ew_{\psi})\ge 0$.
This kind of fidelity witness classifies states as either
(1) the fidelity $ \Tr(\dm_{\prepare}\op{\psi_{\target}}) \le \alpha$; or
(2) the fidelity $ \Tr(\dm_{\prepare}\op{\psi_{\target}}) > \alpha$ implies $\dm\notin\separableset_{\bi}$
\footnote{In other words, the trace distance $\norm{\dm_{\prepare}-\op{\psi_\target}}_1 < \sqrt{1-\alpha}$ because the fidelity and trace distance are related by the inequalities
$1-F\le d_{\tr}(\dm,\dm') \le \sqrt{1-F^2}$ (c.f. \nameref{prm:weak_membership problem_for_separability})}.
For instance, assume the target state is $\ket{\ghz}:=\frac{1}{\sqrt{2}}(\ket{0}^{\otimes n} + \ket{1}^{\otimes n}$),
the maximal overlap between GHZ and bi-separable states is $1/2$,
such that the witness \cref{eq:entanglement_witness} with $\alpha=1/2$ certifies tripartite entanglement
\cite{acinClassificationMixedThreequbit2001}.
We call \cref{eq:entanglement_witness} as projector-based fidelity witness \cite{bourennaneWitnessingMultipartiteEntanglement2004}.
In order to effectively measure a witness in an experiment, it is preferable to decompose the projector term into a sum of locally measurable observables such as 
\footnote{
	$\ew_{\ghz_3} = \frac{1}{8} \qty( 3*III - \px\px\px - \perm(I\pz \pz ) + \perm(XYY))$
	where $\pz \pz I\equiv \pz \otimes\pz \otimes I$ and $\perm(I\pz \pz )\equiv \pz \pz I + \pz I\pz + I\pz \pz$ for readability. 
}.
Meanwhile, for graph states (stabilizer states, i.e., a large class of entanglement states),
a witness can be constructed by very few local measurement settings (LMS) \footnote{For example, the observables $\pz \pz I$, $\pz I\pz$, and $I\pz \pz$ can be measured by one local measurement setting $\pz\pz\pz$.} \cite{tothDetectingGenuineMultipartite2005,tothEntanglementDetectionStabilizer2005,zhouDetectingMultipartiteEntanglement2019}
and implemented in experiments
\cite{luEntanglementStructureEntanglement2018}
\cite{luEntanglementStructureEntanglement2018,zhouSchemeCreateVerify2022},
but non-local measurements are usually required for non-stabilizer cases (e.g., W state) \cite{zhangEfficientEntanglementGeneration2021,zhuMachineLearningDerivedEntanglement2021}.

\section{End-to-end entanglement detection protocol}\label{sec:protocol}
\subsection{Motivation: Beyond fidelity witness}
In most studies of fidelity witness, the robustness measure of a fidelity witness is its tolerance to white noise:
\begin{equation}\label{eq:white_noise}
	\dm
	= (1-p_{\noise}) \op{\psi_{\target}} + p_{\noise} \frac{\identity}{2^{n}}
\end{equation}
where the limit of white noise (i.e., maximal $p_{\noise}$ s.t. $\Tr(\dm\ew_{\psi})<0$) indicates the robustness of the witness.
In general, there are entangled states mixed with large white noise that cannot be detected by conventional methods.
For example, the maximally-entangled Bell state can maximally violate the CHSH inequality, 
but Bell states that mixed with white noise doesn't violate the CHSH inequality when $ 1- 1/ \sqrt{2} < p_{\noise}<2/3 $ despite they are still entangled in this regime.
For 3-qubit GHZ states mixed with white noise, we can analytically compute the white noise threshold for NPT (implies bipartite entanglement):
when $p_{\noise}<0.8$, the states cannot be \nameref{def:bipartite_separable} with respect to any partition (that is \nameref{def:full_entanglement}).
However, the conventional fidelity witness only detects \nameref{def:gme} when $p_{\noise}<4/7$ for GHZ states \cite{guhneEntanglementDetection2009}.
So, it would be practically interesting to have a witness for this white noise regime $p_\noise\in[4/7,0.8)$ 
\footnote{The corresponding white noise regime for W state is $p_\noise\in[8/21,0.791)$} 
that beyond the capability of conventional fidelity witnesses.

Other than white noise, an other typical noise that happens in (photonic) experiments is coherent noise, such as local rotations.
Take $n$-qubit GHZ state as an example, unconscious phase accumulation and 
rotation on the first control qubit can be modeled as 
\cite{zhouEntanglementDetectionCoherent2020}
\begin{equation}
	\ket{\ghz(\phi,\theta)}=
	\cos\theta\ket{0}^{\otimes n}+e^{\ii \phi}\sin\theta\ket{1}^{\otimes n}.
	\label{eq:coherent_noise}
\end{equation}
In a certain noise regime (see Fig. 3 of \cite{zhouEntanglementDetectionCoherent2020}), $\ket{\ghz(\phi,\theta)}$ cannot be detected by conventional fidelity witness because the coherent noise diminishes the fidelity but not change entanglement property.
\begin{figure*}[!ht]
	\centering
	\subfloat{%
		\includegraphics[width=1.0\columnwidth]{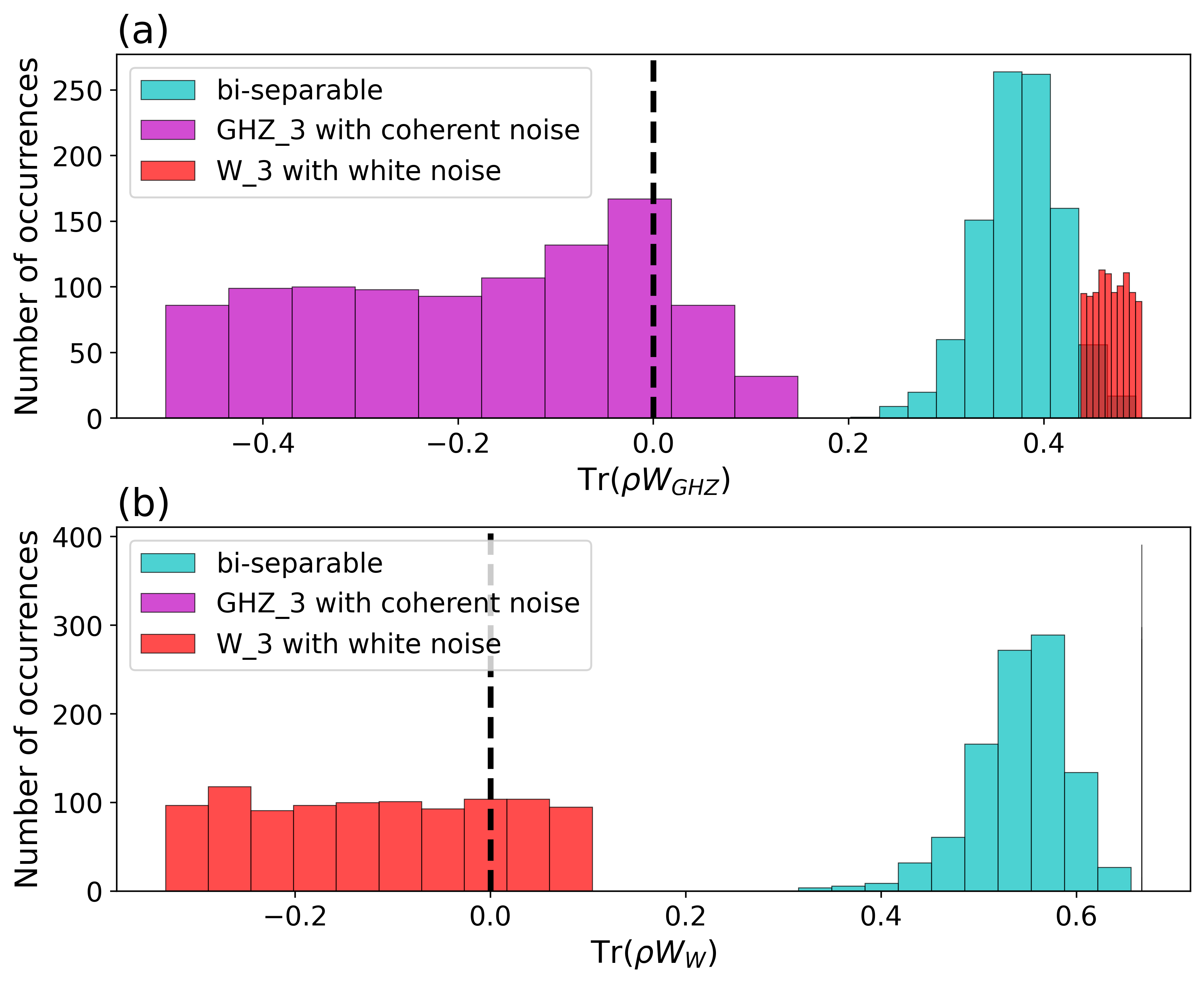}%
	}\hfill
	\subfloat{%
		\includegraphics[width=0.9\columnwidth]{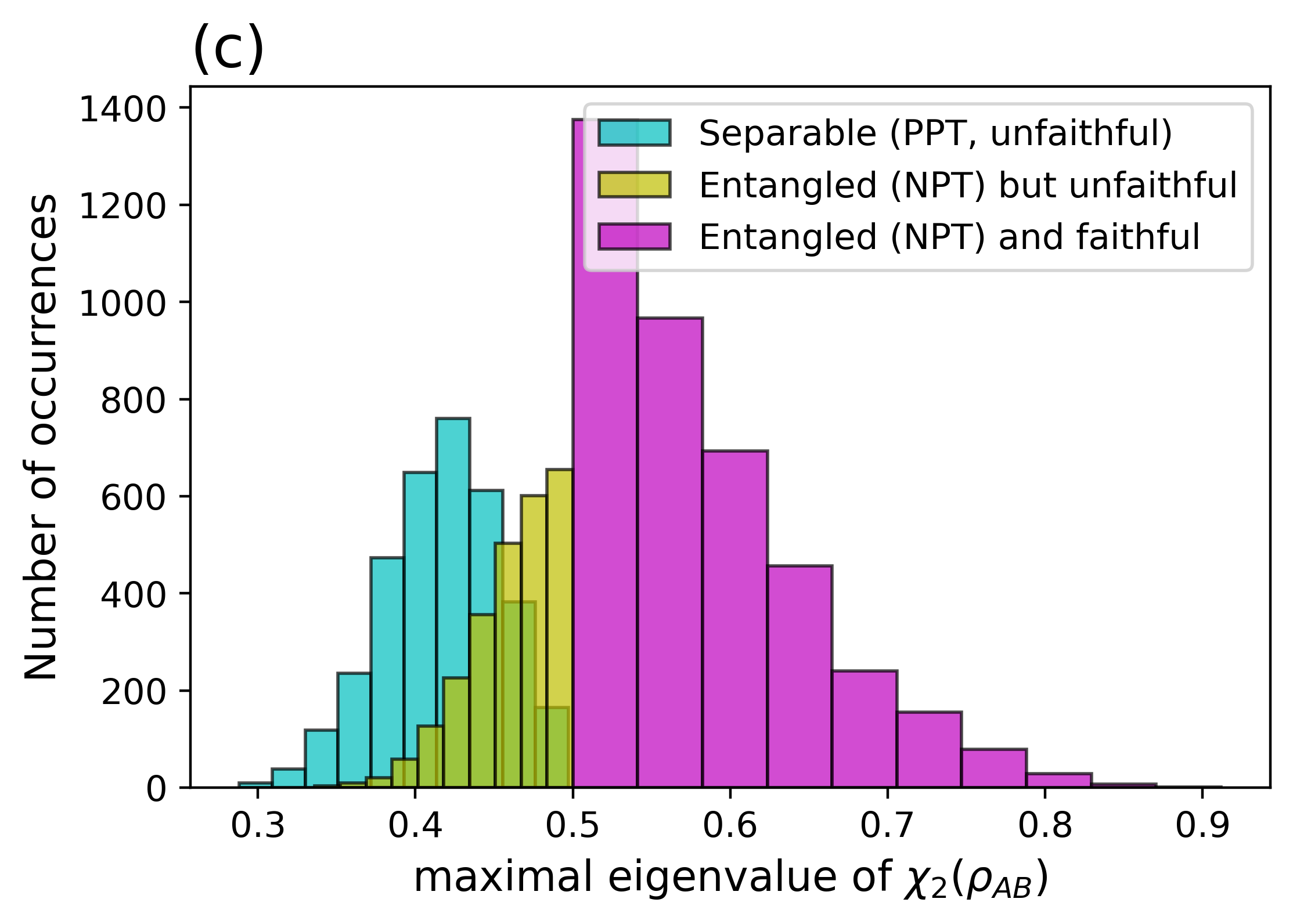}%
	}
	\caption{Examples of the entanglement states cannot be detected by conventional fidelity witnesses. (a) GHZ states with coherent noise sampled with $\theta\in[0,\pi/3]$ and $\phi\in[0.5\pi,0.6\pi]$ cannot be detected by the GHZ projector fidelity witness $W_{\ghz}$ \cref{eq:entanglement_witness}. Entangled states should be on the left of the dashed vertical line, i.e., have a negative expectation value of the witnesses $\Tr(\dm\ew)$. (b) Similarly, full entanglement of W states with large white noise $p_{\noise}\in[8/21,0.5]$ cannot be detected by $W_{\text{w}}$. And we can see W states with white noise has $\Tr(\dm_{\text{W}}\ew_{\ghz})>0$, vice versa. (c) Unfaithfulness of 2-qubit states: $10^4$ randomly sampled 2-qubit states are categorized according to the minimal eigenvalue of partial transpose $\dm_{AB}^{\T_A}$ and the maximal eigenvalue of $\chi_2(\dm_{AB})$.}
	\label{fig:conventional_witness}
\end{figure*}

To formally characterize the cases beyond fidelity witness, Weilenmann et. al \cite{weilenmannEntanglementDetectionMeasuring2020} \cite{huOptimizedDetectionHighDimensional2021} coined the term \emph{unfaithful states} 
which systematically analyzes a 2-qudit entangled state mixed with white noise that cannot be detected by fidelity witness.
They found that for $d \ge 3$ that almost all states in the Hilbert space are unfaithful. 
Subsequently, G\"{u}the et. al \cite{guhneGeometryFaithfulEntanglement2021} \cite{riccardiExploringRelationshipFaithfulness2021} gave a formal definition: 
	a 2-qudit state $\dm_{AB}$ is faithful if and only if there are local unitary transformations $\U_A$ and $\U_B$ such that
	$\expval{\U_A\otimes\U_B \dm_{AB} \U_A^\dagger \otimes\U_B^\dagger}{\phi^+}> \frac{1}{d}$.
Consequently, they found a necessary and sufficient condition for 2-qubit unfaithfulness: 
a 2-qubit state $\dm_{AB}$ is faithful if and only if the maximal eigenvalue of
\begin{equation}
	\mathcal{X}_2( \dm_{AB})=\rho_{AB}-\frac{1}{2}(\dm_{A}\otimes I + I \otimes \dm_{B})+\frac{1}{2} I \otimes I
	\label{eq:unfaithful_2qubit}
\end{equation}
is larger than 1/2.
We can see in (c) of \cref{fig:conventional_witness}, even for 2-qubit states, nonnegligible portion of randomly sampled states are unfaithful but still entangled (NPT).

Although there are variants of witness, such as nonlinear witness \cite{guhneNonlinearEntanglementWitnesses2006} and post-processing \cite{zhanDetectingEntanglementUnfaithful2021}, designed to remedy the shortcomings of conventional fidelity witness respectively, 
it would be meaningful in practice to find a generic method to construct witnesses (classifiers) for \nameref{prm:entanglement_detection}.
Machine learning techniques suit the needs well because supervised learning can be regarded as a powerful nonlinear post-processing tool.

\subsection{Training a generic witness via kernel SVM}
One basic task in classical machine learning (ML) is binary classification,
such as cat/dog image classification. 
In this case, the input to a ML algorithm is a (training) dataset $\qty{(\vbx^{(i)},y^{(i)})}_{i=1}^m$ consists of $m$ data points, 
where each data point is a pair of feature vector $\vbx\in \realnumber^d$ of $d$ features and its label $y\in\qty{-1,1}$.
For example, the feature $\vbx$ of an image is a flattened vector of all pixel values and the label $y=-1$ for \textsc{cat} images ($1$ for \textsc{dog}).
It is clear that \nameref{prm:separability} or \nameref{prm:entanglement_detection} problem are exactly such binary classification problems where each quantum state has a binary label, such as either `\entangled' or `\separable'.
The features $\vbx$ of a quantum state $\dm$ can be the entries of its density matrix, or more realistically, the expectation values of selected observables.

With the surge of research on ML, 
classification tasks related to entanglement have been performed by ML algorithms.
Lu et. al \cite{luSeparabilityEntanglementClassifierMachine2018} 
trained a (universal) \nameref{prm:separability} classifier by classical neural network
where features of $\vbx$ are the entries of density matrices.
For the similar purpose, Ma and Yung \cite{maTransformingBellInequalities2018} generalized Bell inequalities to a Bell-like ansatz $\ew_{\ml}:=\vbw_{\ml}\cdot\vb{\pob}_{\bellineq}$ where the optimal weights $\vbw_{\ml}$ are obtained via optimizing a neural network.
And they found the tomographic ansatz
\begin{equation}
	\Tr(\dm\ew_{\ml}) \equiv\expval{\ew_{\ml}}  \equiv
	\vbw_{\ml} \cdot \expval{\vb{\pob}_{\sigma}}, 
	\label{eq:tomographic_ansatz}
\end{equation}
not only has better performance than the Bell-like ansatz, 
also required \cite{luTomographyNecessaryUniversal2016} for training a universal \nameref{prm:separability} classifier,
where the feature vector $\vbx_{\dm,\bmsigma}:=\expval{\vb{\pob}_{\sigma}}$ denotes the expectations of all $4^n$ Pauli observables
\footnote{
	Denote $\pob_{\sigma}\in \qty{I,X,Y,Z}^{\otimes n}$ for a Pauli observable.
	Denote $\vbx_{\dm,\bmsigma}:=(\Tr(\dm\pob_{\sigma_1}),\dots,\Tr(\dm\pob_{\sigma_M}))$ for a vector of expectations of $M$ Pauli observables $\bmsigma\subseteq \qty{I,X,Y,Z}^n$ measured on $\dm$. 
}.
It is worth noting that training such a universal classifier for high-dimensional systems needs a large training dataset and long time if the gap between two state sets is small.

\begin{figure*}[!ht]
	\centering
		\centering
		\includegraphics[width=.56\linewidth]{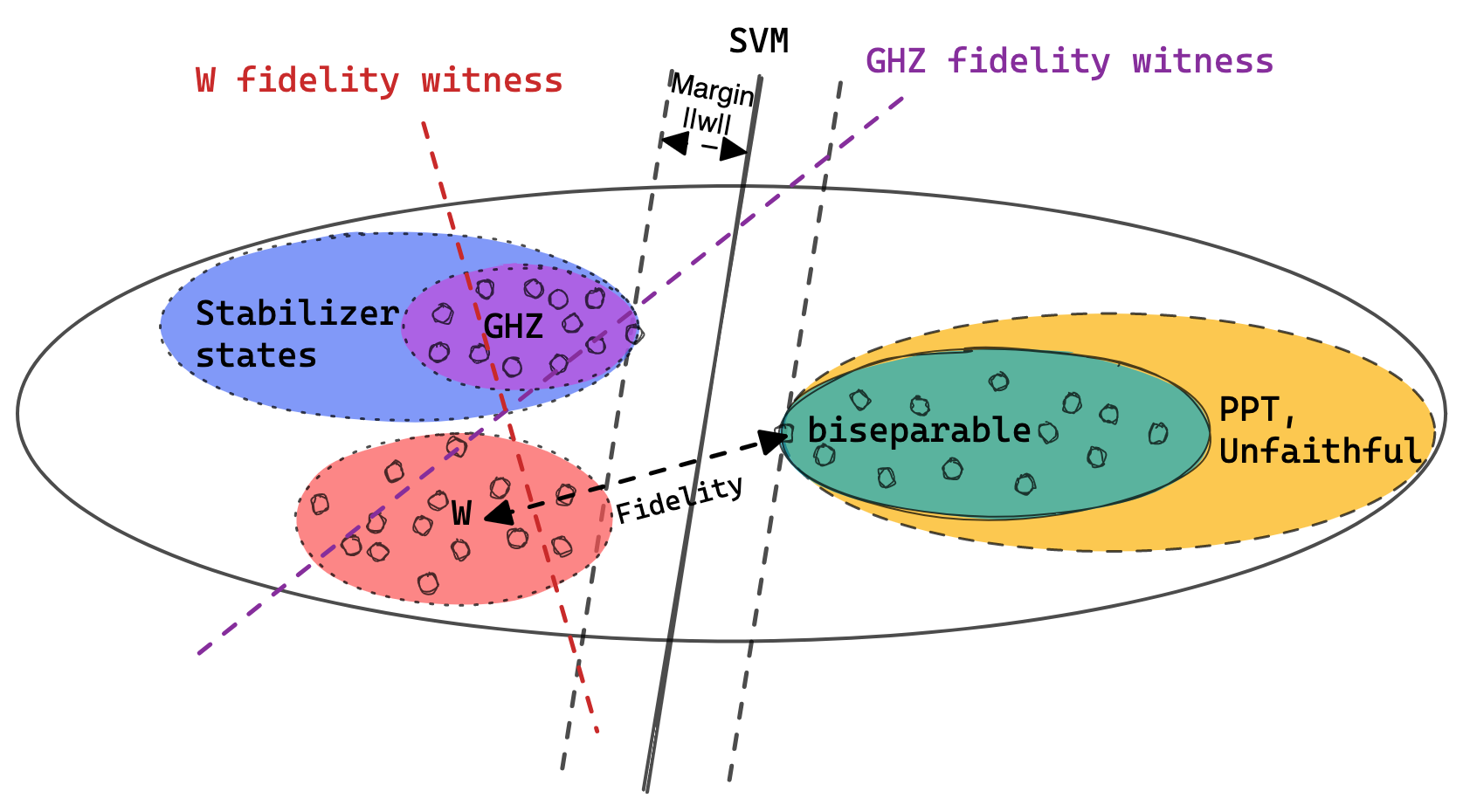}
	\caption{Schematic diagram for different entanglement detection methods: the colored ellipses with dotted boundary indicate the vicinity (white noise) of certain entangled states such as GHZ, and W states. Conventional fidelity witnesses for different states are depicted by colored dashed lines (hyperplanes in feature space). An entangled state with large white noise or coherent noise (local rotation depicted by a curve) cannot be detected by conventional fidelity witnesses. SVM without a nonlinear kernel is a hyperplane separating two sets of colored dots (synthetic dataset). The data points on the boundaries (dashed black lines) are called support vectors. The distance between the SVM hyperplane and boundary is the margin to be minimized via optimization. The PPT criterion is a nonlinear but one-side classifier without prior knowledge. The circles in each ellipses indicate the sampled states for training.}
	\label{fig:entangle_schematic}
\end{figure*}

In this paper, we focus on solving the \nameref{prm:entanglement_detection} problem with training data.
In other words, we derive the entanglement witness (classifier) for certain target states with desired entanglement structure by fitting a synthetic dataset.
\begin{problem}[learning an entanglement witness]\label{prm:learn_witness}
	\;
	\begin{itemize}
		\item \textbf{Input}: a dataset $\qty{\qty(\dm^{(i)},y^{(i)})}$ consist of entangled states $\dm$ around $\ket{\psi_{\target}}$ with label $y=-1$ and randomly sampled bi-separable states with label $1$.
		\item \textbf{Output}: a classifier $f(\vbx_{\dm,\tilde{\bmsigma}})$ with high training accuracy where $\tilde{\bmsigma}$ is a subset of all Pauli observables and $\vbx_{\dm,\tilde{\bmsigma}}$ is a vector of corresponding expectation values.
	\end{itemize}
\end{problem}

This problem has also been studied by classical ML \cite{zhuMachineLearningDerivedEntanglement2021}  \cite{vintskevichClassificationFourqubitEntangled2022}, 
but by a technique different from Neural Network (NN), called Support Vector Machine (SVM)  \cite{cortesSupportvectorNetworks1995}.
A classification task performed by SVM can be formulated as a convex optimization problem:
find a hyperplane parametrized by $(\vbw,b)$  in a feature space (a linear function $f$) that maximizes the margin between two decision boundaries subject to the constraint that two types of data points are separated (on the two sides of the hyperplane, see \cref{fig:entangle_schematic})
\begin{equation}
	\max_{\vb{w}}
	\norm{\vb{w}}_2^2 \;
	\text{ s.t. }
	\forall i,\; y^{(i)}\cdot (\vb{w}\cdot\vbx^{(i)}+b)\ge 1.
	\label{eq:svm_opt}
\end{equation}
where $\vbw$ is the (not necessarily normalized) normal vector to the hyperplane and $b$ is a bias term similar to $\alpha$ in \cref{eq:entanglement_witness}.
Therefore, the predicted label is given by the sign of the inner product (projection) between the hyperplane and the feature vector $\vbx$, i.e., $y=f(\vbx)=\text{sign}(\vbw\cdot\vbx+b)$ (c.f. \cref{eq:witness} and \cref{eq:tomographic_ansatz}).
Geometrically, both SVM witness and conventional fidelity witness
are hyperplanes in feature spaces,
but the SVM witness is more flexible because the classifier $(\vbw,b)$ can be numerically derived through optimization for any generic target state.
And it can only require local Pauli observables (measurements) $\pob_{\sigma}$ that is feasible in most experiments, even when the target state is a non-stabilizer state.

The SVM allows for the programmatic elimination of features \cite{guyonGeneSelectionCancer2002}, i.e., reducing the cost of experimental measurements (samples).
We start with the feature vector of all $k$-local Pauli observables,
then we randomly eliminate one feature such that the training accuracy remains high enough with the new feature vector $\tilde{\vbx}$.
By repeating this procedure, we obtain a classifier $f(\vbx_{\dm,\tilde{\bmsigma}})$, 
where $\abs{\tilde{\bmsigma}}=M$ is the minimal number of Pauli observables required for classification.
The algorithm is summarized in Algorithm. \ref{alg:classical_learning}.
\begin{algorithm}[H]
    \DontPrintSemicolon
    \SetKwInOut{Input}{input}
    \SetKwInOut{Output}{output}
    \Input{dataset $\qty{\qty(\dm^{(i)},y^{(i)})}^m_{i=1}$, minimal number of features: $M$, and tolerance $\epsilon$}
    \Output{a classifier $f(\vbx_{\dm,\tilde{\bmsigma}})$ }
    \BlankLine
	$\vbx^{(i)}:=\Tr(\dm^{(i)}\pob_{\bmsigma}), \forall i$ \tcp*{evaluate all $k$-local Pauli observables and shuffle}
	\While{\textup{accuracy $< \epsilon$ or} $len(\vbx)>M$ } {
		\For{$j \textup{ in  range}(len(\vbx)) $} {
			\tcc{eliminate $j$-th feature}
			$\forall i$, let $\tilde{\vbx}^{(i)}$ be $\vbx^{(i)}$ without the $j$-th feature\\
			\tcc{Train SVM with the new feature vectors}
			accuracy, classifier = SVM($\qty{\qty(\tilde{\vbx}^{(i)},y^{(i)})}^m_i$)
			\If{\textup{accuracy $\ge \epsilon$ } } {
				$\vbx^{(i)}:=\tilde{\vbx}^{(i)}$ and then \textbf{break}
			}\ElseIf{\textup{accuracy $<\epsilon$  and} $j=len(\vbx)$} {
			\tcc{If cannot find a classifier with fewer features, then output the last classifier with high accuracy}
				\Return a classifier $f(\vbx_{\dm,\tilde{\bmsigma}})$
			}
		}
	} \Return a classifier $f(\vbx_{\dm,\tilde{\bmsigma}})$ with $\abs{\tilde{\bmsigma}}=M$
    \caption{Train a witness via kernel SVM }
    \label{alg:classical_learning}
\end{algorithm}
\begin{figure*}[!ht]
	\centering
	\subfloat{%
		\includegraphics[width=.9\columnwidth]{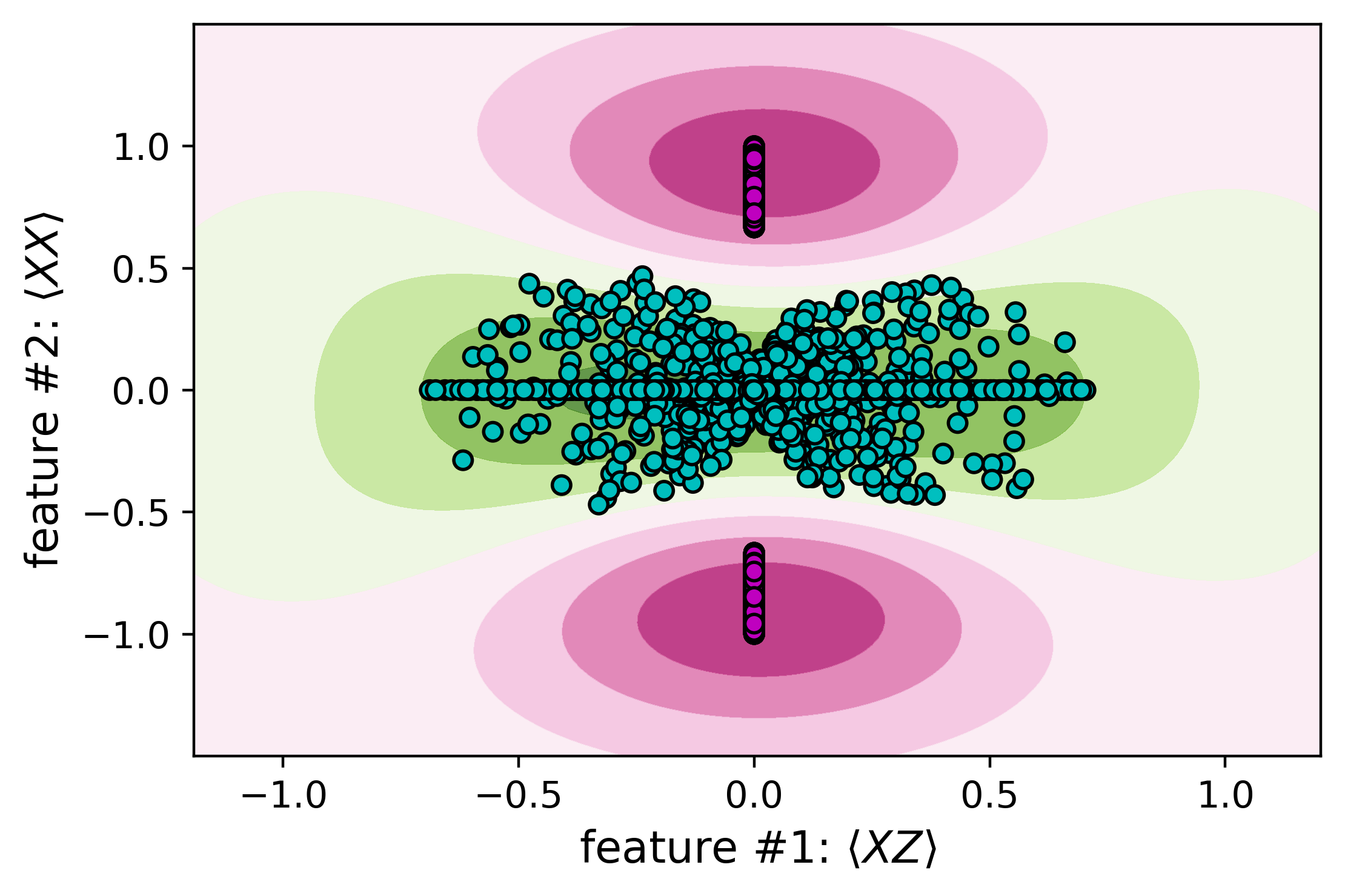}
	}\hfill
	\subfloat{%
		\includegraphics[width=.9\columnwidth]{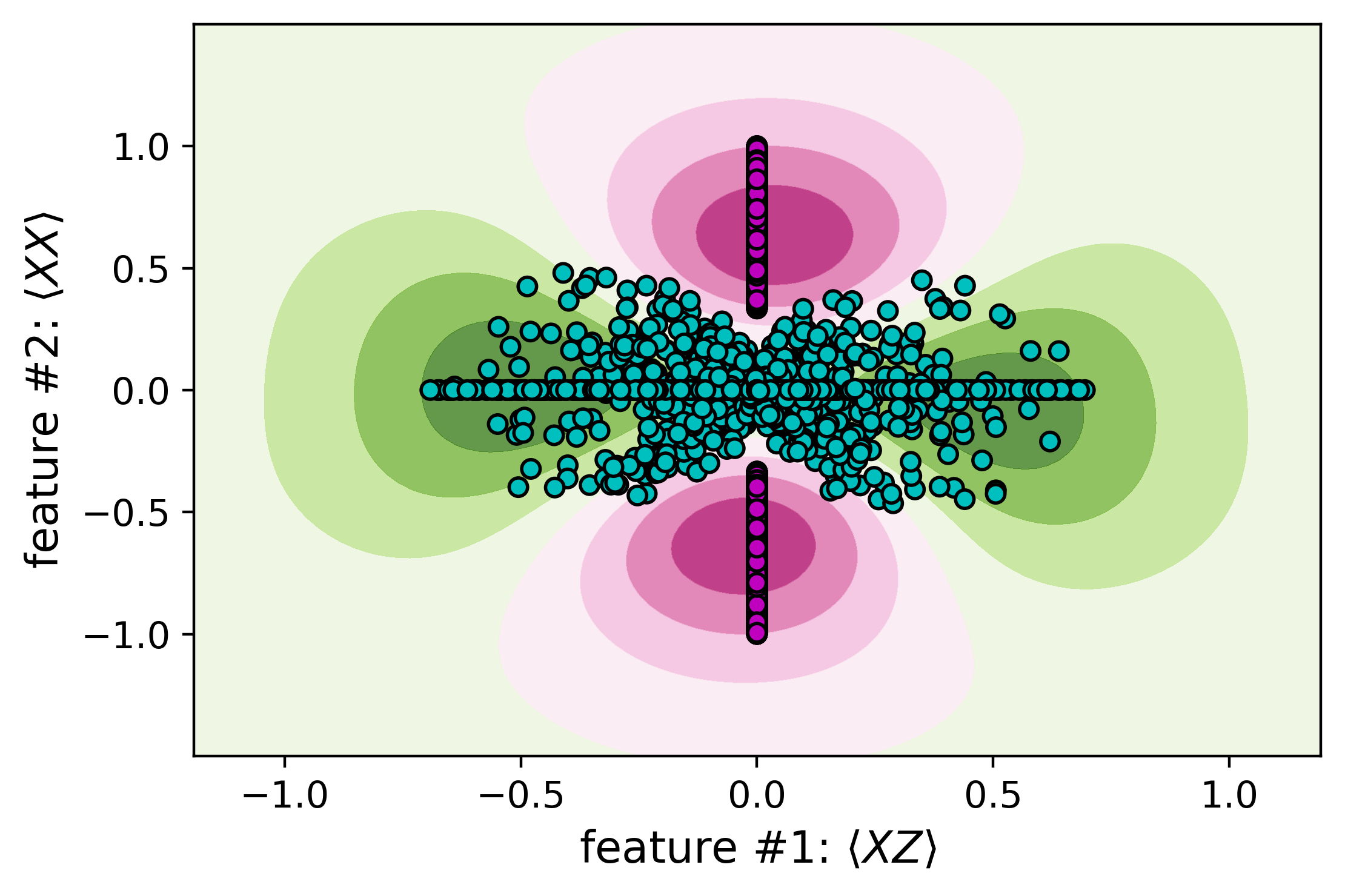}
	}
	\caption{The two-dimensional embedding (a low-dimensional feature space $\expval{XZ}$ VS $\expval{XX}$) of 2-qubit states: green dots represent randomly sampled separable states, while pink ones represent entangled Bell states mixed with white noise in the range (left figure) $p_{\noise}\in[0,1/3]$ and (right figure) $p_{\noise}\in[0,2/3]$. The colored shade indicates the nonlinear decision boundary of the RBF kernel SVM classifier. When the white noise is larger, the gap between two sets of data points is smaller such that training a classifier becomes harder.}
	\label{fig:feature_space}
\end{figure*}
A key drawback of both conventional witnesses and SVM is their linearity because many real-world datasets are not linearly-separable in a low-dimensional feature space.
Despite the nonlinear witness \cite{guhneNonlinearEntanglementWitnesses2006} proposed, its experimental implementation is more challenging than linear ones.
The good news is,
within the framework of SVM, non-linearity can be easily achieved by the so-called kernel method \cite{hofmannKernelMethodsMachine2008}.
The main idea is mapping the features $\vbx$ to a higher dimensional space via a feature map $\phi(\vbx)$ such that they can be linearly separated in the high-dimensional feature space.
The kernel function $\kernel(\vbx,\vbx'):\mathcal{X}\times \mathcal{X} \to \realnumber$ measures the similarity between two input data points in the high-dimensional feature space because a kernel can be written as an inner product $\langle \phi(\vbx),\phi(\vbx') \rangle$.
The commonly used kernel is the radial basis function (RBF) kernel which a Gaussian function
$\kernel_{\text{rbf}}(\vbx,\vbx') := \exp(-\gamma\norm{\vbx-\vbx'}^2_2)$ with $l_2$ Euclidean norm and a parameter $\gamma$.
Since the RBF kernel SVM is convex, the optimal classifier function will be found if it exists for the input dataset.
The power of the kernel method can be clearly observed in \cref{fig:feature_space} that two kinds of data points are clearly classified by a nonlinear (RBF kernel) SVM classifier, though it is not linearly separable in this 2-dimensional space.

\begin{table}[!ht]
	\centering
	\begin{tabular}{c|c|c|c}
		Witnesses & \# observables & weights & comment \\
		\hline
		Conventional fidelity   & few LMS & fixed & one-side  \\  
		\textbf{SVM (kernel) } &  $\ll 4^n-1$ & trained & flexible \\  
		Tomographic (NN)  & $4^n-1$ & trained & universal \\  
		\hline
	\end{tabular}
	\caption{Comparison of conventional fidelity witness, tomographic classifier, and SVM witness.}
	\label{tab:comparison}
\end{table}
We compare the characters of different kinds of witnesses in \cref{tab:comparison}. 
The conventional fidelity witness only need few local measurement settings for stabilizer states, but it is a one-side test. 
The tomographic witness trained by NN only need the promise that there is a gap between entangled and separable states (almost universal), but it requires complete information of a state ($4^n-1$ features).
Between these two cases, the SVM witness has stronger classification capability than conventional fidelity witnesses and do not need as many classical features as the tomographic witness.
However, these prior ML witnesses only consider the robustness to white noise and cannot be directly applied to experiments.
In the numerical simulation, we can efficiently evaluate classical features by direct calculation, 
but in actual experiments, entries of a density matrix are not explicitly known.
Instead, we need to estimate observables (classical features) by repeat measurements, which we are going to discuss in next section.

\subsection{Sample-efficient expectation estimation methods}\label{sec:estimation}
The brute force approach to fully characterize a state in an experiment is quantum state tomography \cite{altepeterPhotonicStateTomography2005}
\footnote{Quantum state tomography refers to the task of recovering the density matrix of an unknown $D$-dimensional state $\dm$ within error tolerance $\epsilon$, 
given the ability to prepare and measure copies of $\dm$.}.
With a recovered density matrix, we can directly calculate classical features or separability measures,
but full tomography is experimentally demanding.
Even adaptive or collective measurements (and post-processing) allowed 
\footnote{Adaptive measurements are the intermediate between independent measurements and collective (entangled) measurements, in which the copies of $\dm$ are measured individually, but the choice of measurement basis can change in response to earlier measurements.},
rigorous analysis \cite{haahSampleoptimalTomographyQuantum2017} \cite{odonnellEfficientQuantumTomography2016} showed that 
$\Omega(D^2/\epsilon^2)$ measurements (copies)  are required for recovering a $D\times D$ density matrix
with error tolerance $\epsilon$ measured by trace distance.
Now that full tomography is intractable for large systems, a workaround is to extract partial information about a state without fully recovering it:
\begin{problem}[shadow tomography]\label{prm:shadow_tomography}
	Given $m$ copies (samples) of an unknown $D$-dimensional state and $M$ known 2-outcome measurements $\qty{E_1,\dots,E_M}$,
	to estimate $\forall i,\Tr(\dm E_i)$ within additive error $\epsilon$ with success probability at least $1-\delta$.
\end{problem}
Since shadow tomography can be implemented with $\tilde{\bigO}(\log^4 M\cdot \log D\cdot \log 1/\delta \cdot\epsilon^{-4})$ copies \footnote{The notation $\tilde{\bigO}$ hides a polylog factor. A full tomography requires estimate $D^2$ measurements (observables) with additive error $\epsilon\ll 1/D$ for all $E_i$, so the sample complexity of shadow tomography is compatible with lower bounds of full quantum state tomography.} \cite{aaronsonShadowTomographyQuantum2018},
we can estimate $M$ classical features (Pauli observables) for $f(\vbx_{\dm,\tilde{\bmsigma}})$ in a samples-efficient manner.
However, Aaronson's shadow tomography procedure is very demanding in terms of quantum hardware (in the collective preparation and measurement on $\dm^{\otimes m}$).
To be more feasible for current experiments, Huang et. al \cite{huangPredictingManyProperties2020} introduced a classical shadow (CS) scheme which we apply in our protocol.

\begin{algorithm}[H]
    \DontPrintSemicolon
    \SetKwInOut{Input}{input}
    \SetKwInOut{Output}{output}
    \Input{$R$ copies of $\dm$ and selected observables $\pob_{\tilde{\bmsigma}}$}
    \Output{estimation of $\vbx_{\dm,\tilde{\bmsigma}}:=\Tr(\dm\pob_{\tilde{\bmsigma}})$}
    \BlankLine
	Sample $R$ Pauli measurements $P\in\qty{X,Y,Z}^{\otimes n}$ \\
    \For{ $i = 1,2, \ldots, R$} {
		\tcp{apply single-copy measurement $P$ to a copy $\dm$} 
        $\dm\mapsto \U\dm \U^\dagger\mapsto \ket{\vb{b}}$ with $b_j\in \qty{0,1},\forall j\in[n]$ \\
		\tcp{inverse channel $\mathcal{M}^{-1}(\dm')=(3\dm'-I)$}
		$\dm_{\cs}^{(i)}=\bigotimes_j^n \qty(3U_j^\dagger \op{b_j} U_j-I)$ 
    }
    $\text{CS}(\dm,R)=\qty{\dm_{\cs}^{(1)},\dots,\dm_{\cs}^{(R)}}$ \tcp*{classical shadow}
	\tcp{estimate features for SVM from classical shadow}
	\Return  $\vbx_{\dm,\tilde{\bmsigma}}=\textsc{Expectation}(\text{CS}(\dm,R)O_{\tilde{\bmsigma}})$
    \caption{Estimate Pauli observables (features) by randomized classical shadow}
    \label{alg:classical_shadow}
\end{algorithm}
The classical shadow of a state $\dm$ (a set of snapshots $\dm_{\cs}$) is a succinct classical description of a state $\dm$, which can be used to estimate the expectations of a sef of observables with a reasonably small number of copies of $\dm$.
To construct the randomized classical shadow, we first need to uniformly sample $R$ Pauli measurements $P\in\qty{X,Y,Z}^{\otimes n}$ (assume the state $\dm$ of $n$ qubits).
Then, we apply single-copy measurement $P$ to a copy of $\dm$, i.e., each measurement measures all qubits in Pauli $X$, $Y$, or $Z$-basis according to $P$.
Specifically, we apply the transformation $\dm \mapsto \U \dm \U^\dagger$ where $U P U^\dagger = \Sigma$ is the eigendecomposition of $P$ and then measure this rotated stated in computational basis (collapse to $\ket{\vb{b}}\in\qty{\ket{0},\ket{1}}^{\otimes n}$).
A snapshot $\dm_{\cs}=\mathcal{M}^{-1}(U^\dagger\op{\vb{b}}U)$ can be constructed by taking the inverse of the quantum depolarizing channel $\mathcal{M}$.
By repeating this procedure $R$ times, we have $R$ snapshots of $\dm$ to estimate expectation values of a set of Pauli observables by an empirical average over snapshots, 
i.e., $o = \Tr(O \dm_{\cs})$ obeys $\expectation[o] =\Tr(O \dm)$.
The algorithm is summarized in Algorithm. \ref{alg:classical_shadow}.

Surprisingly, by rigorous proof, the size of the classical shadow scales $\bigO(\log(M) 3^k/\epsilon^2)$ to approximate $M$ $k$-local Pauli observables with error tolerance $\epsilon$ \cite{huangPredictingManyProperties2020},
so this scheme has an advantage for small $k$ and large $M$ cases (many very local observables).
For different purposes, there are several variants of the classical shadow method \cite{hadfieldMeasurementsQuantumHamiltonians2022, huangEfficientEstimationPauli2021, chenRobustShadowEstimation2021}.
The derandomized version \cite{huangEfficientEstimationPauli2021} is the refinement of the original randomized protocol which provides better performance for $k$-local observables. 
The core idea of the derandomized version is to sample more global Pauli measurements that are compatible with $k$-local Pauli observables to be estimated.
This procedure is the interpolation between completely randomized measurements (the original classical shadow) and independent estimation (good for predicting a few global observables).
From the perspective of a conventional entanglement witness, the classical shadow method finds an effective local measurement setting for a generic set of $k$-local Pauli observables.
In addition, the entanglement detection by estimating $p_3$-PPT with classical shadow \cite{elbenMixedstateEntanglementLocal2020} and comparison of classical shadow variants \cite{zhangExperimentalQuantumState2021} has been done experimentally.

\section{Numerical simulation and Discussion}\label{sec:numerical_simulation}

In numerical simulation, we generate quantum state samples and manipulate quantum states by QuTiP Python library \cite{johanssonQuTiPPythonFramework2013,liPulselevelNoisyQuantum2022}.
Multi-partite entangled states including Bell states, GHZ states with coherent noise \cref{eq:coherent_noise} and W states with white noise \cref{eq:white_noise} are generated by calling functions provided by QuTiP.
The noise parameters $(\phi,\theta,p_{\noise})$ are uniformly sampled from certain ranges.
In contrast to entangled states, we generate random separable states for different number of qubits by tensoring random density matrices of subsystems.
For example,
there are three different partitions $\dm_1\otimes \dm_{23}$, $\dm_{12}\otimes \dm_{3}$, and $\dm_2\otimes \dm_{13}$ for 3-qubit bi-separable states.
It is not necessary to prepare the (mixed) separable states as convex combination of separable states with different partitions 
because SVM can correctly classifiy a mixture if it can classify each case.

\begin{figure}[!ht]
	\centering
		\includegraphics[width=.9\linewidth]{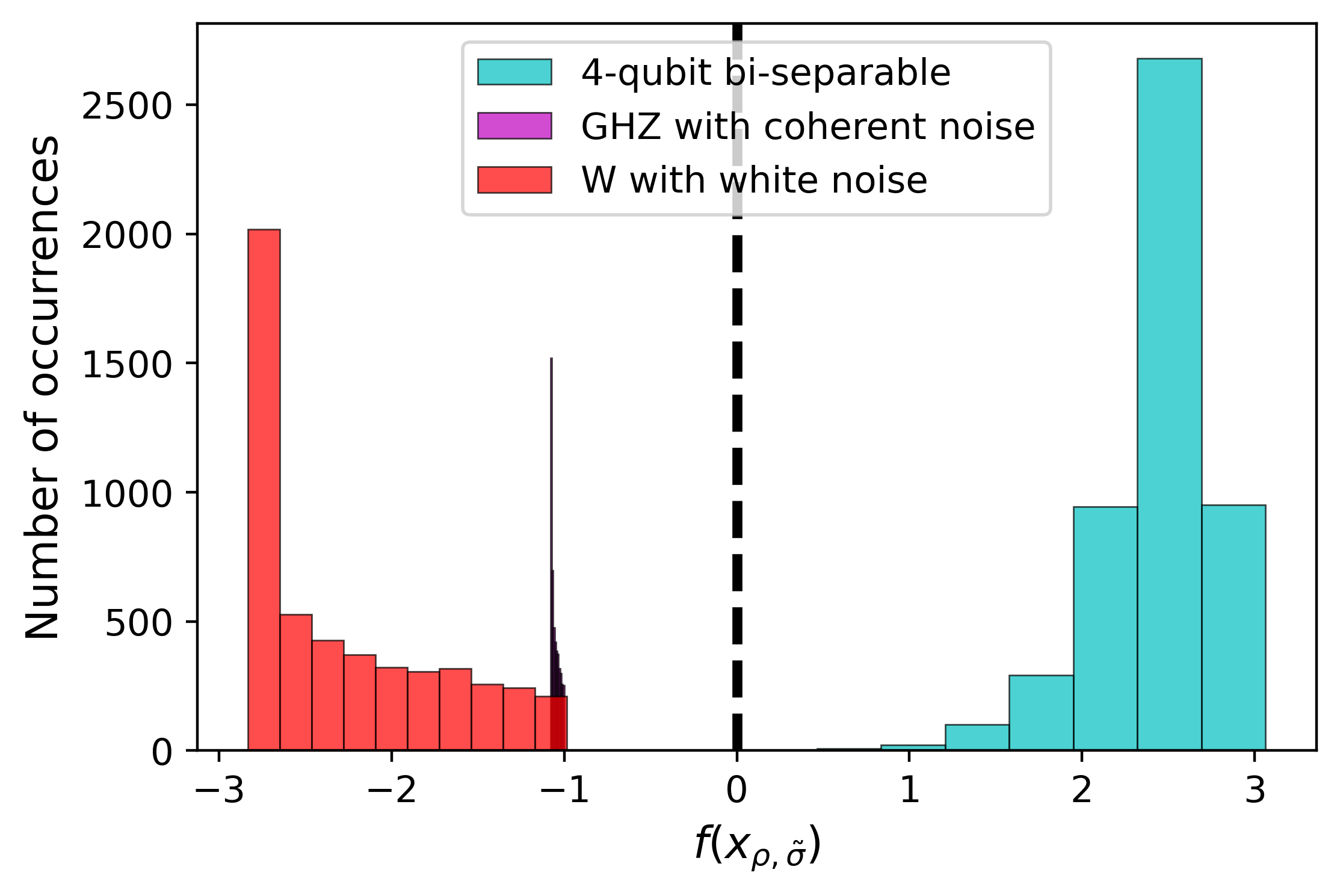}
	\caption{The 4-qubit GHZ state with coherence noise $\theta\in[0,\pi/3], \phi\in[0,0.6\pi]$ and white noise $p_{\noise}\in[0,0.1]$, and the 4-qubit W state with white noise $p_{\noise}\in[0,0.5]$ can be classified by the RBF kernel SVM classifier with high accuracy.}
	\label{fig:ml_compare}
\end{figure}

For the machine learning part, we make use of scikit-learning Python package \cite{pedregosaScikitlearnMachineLearning2011} to train SVM with RBF kernel.
It has been shown in \cref{fig:conventional_witness} that conventional fidelity witnesses cannot correctly classify when 3-qubit GHZ states with coherent noises $\theta=\pi/3,\phi\in[0.5\pi,0.6\pi]$ and W states mixed with white noise $p_{\noise}\in[8/21,0.5]$.
In contrast, the SVM classifier can detect the entanglement of 4-qubit (more challenging than 3-qubit case) GHZ state with coherence noise $\theta\in[0,\pi/3], \phi\in[0,0.6\pi]$ (even mixed with white noise $p_\noise \in [0,0.1]$) and the 4-qubit W state with white noise $p_{\noise}\in[0,0.5]$ ($4/15$ is maximal white noise tolerance of 4-qubit W projector fidelity witness), with high accuracy (see \cref{fig:ml_compare}).
To train this 4-qubit SVM classifier with accuracy $0.999$, we generate $10^4$ states for each kind of states: noisy GHZ, W states with noise parameters uniformly sampled, 
bi-separable states $\dm_{1}\otimes \dm_{234}$ and $\dm_{12}\otimes \dm_{34}$.
\begin{figure}[!ht]
	\centering
	\includegraphics[width=.9\linewidth]{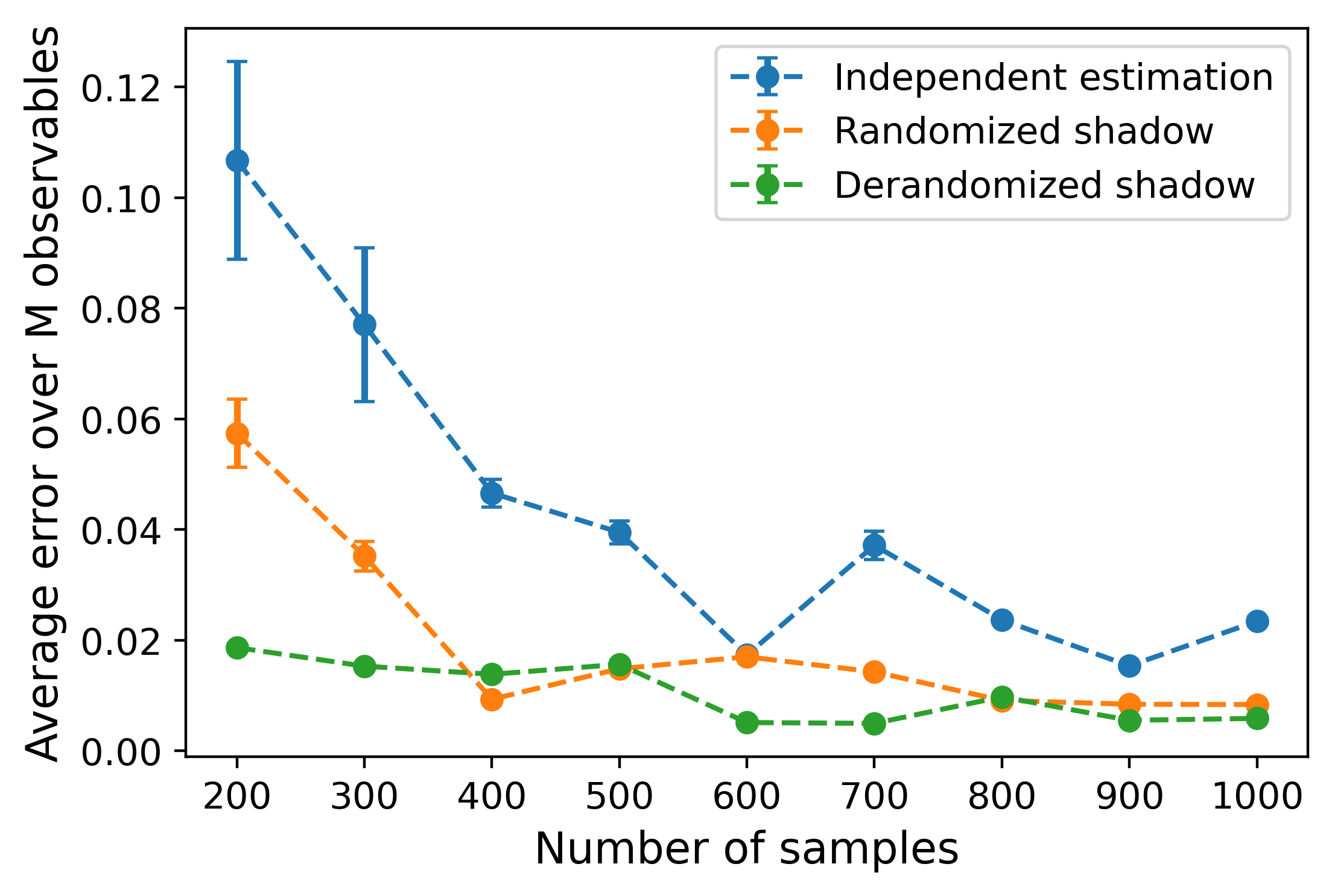}
	\caption{Average error of estimating expectation values of 22 two-local four-qubit Pauli observables $\sum_i (o_i-\Tr(\ob_i\dm))^2/22$ by three different estimation methods VS the number of samples (the error bar indicates the variance of estimation errors over 22 observables). The blue line represents estimating each Pauli observable by repeating measurements independently. The orange line represents the estimation by the randomized classical shadow, while the green one represents the derandomized version of the classical shadow.}
	\label{fig:shadow}
\end{figure}

By programmatic elimination of features, one set of features (i.e., 4 two-local Pauli observables) found by the kernel SVM is
$\vbx=\expval{XIIX},\expval{YIIZ},\expval{IIZZ},\expval{ZXII})$.
You may have noticed that we only consider two partitions in a bi-separable state.
Fortunately, by the symmetry of GHZ and W states, we can exchange a set of qubits positions $[(1,2),(1,3),(1,4),(2,3),(2,4),(3,4)]$ to get all partitions of biseparable states. 
So, there are in total $4*7=28$ features to be estimated (actually 22 due to duplication).
To compare the performance of different estimation schemes, the average error of estimation over observables VS the number of samples is plotted in \cref{fig:shadow}
\footnote{The open-source code for classical shadow with the code from \url{https://github.com/hsinyuan-huang/predicting-quantum-properties}}.
The derandomized version outperforms randomized shadow for a small number of samples and is significantly better than independent estimation (smaller average error and variance).
Notably, the classical shadow estimation of comparable size has been implemented in photonic experiments \cite{zhangExperimentalQuantumState2021}.

In conclusion,
our protocol is flexible and sample-efficient in detecting entanglement in the vicinity of entangled states.
A generic entanglement classifier for a target state that can be viewed as a nonlinear entanglement witness is obtained by training a kernel SVM.
This protocol can be efficiently implemented on current experiment devices because the number of features has been optimized through SVM and efficient local measurement settings are found by the derandomized classical shadow scheme.
Meanwhile, there are also several potential directions for future research:
(1) It is of theoretic interest to find rigorous proof for the dataset size and number of features (required for high training accuracy) scaling with the system size;
(2) It is meaningful to test more kernels, such as graph kernel \cite{vishwanathanGraphKernels2010}, shadow kernel \cite{huangProvablyEfficientMachine2022}, and neural tangent kernel \cite{jacotNeuralTangentKernel2020}, for better performance of the kernel SVM.
And quantum kernel methods \cite{schuldQuantumMachineLearning2019,schuldSupervisedQuantumMachine2021,liuRigorousRobustQuantum2021} might provide advantages over classical counterparts.
(3) 
The task of estimating expectation values can also be achieved efficiently by classical \cite{gaoEfficientRepresentationQuantum2017,torlaiManybodyQuantumState2018,zhuFlexibleLearningQuantum2022} and quantum machine learning \cite{huangPowerDataQuantum2021,huangProvablyEfficientMachine2022}.
Huang et. al rigorously showed that, 
for achieving accurate prediction on all $4^n-1$ Pauli observables 
the exponential quantum advantage over classical ML is possible
\cite{huangInformationtheoreticBoundsQuantum2021}.
Training a more powerful (almost universal) classifier with all Pauli observables as features might be interesting for practice.

We thank HKU/CS Summer Research Internship Program for providing the opportunity and funding for this project.

\bibliographystyle{apsrev4-2}
\bibliography{ref}

\onecolumngrid
\appendix

\end{document}